**GEMINI: integrative exploration of genetic variation and genome annotations.**

Uma Paila[1], Brad Chapman[2], Rory Kirchner[2], and Aaron Quinlan[1]*

[1] Department of Public Health Sciences and Center for Public Health Genomics, University of Virginia, Charlottesville, VA, USA
[2] School of Public Health, Harvard University, Boston, MA, USA

* to whom correspondence should be addressed.

## Abstract

Modern DNA sequencing technologies enable geneticists to rapidly identify genetic variation among many human genomes. However, isolating the minority of variants underlying disease remains an important, yet formidable challenge for medical genetics. We have developed GEMINI (GEnome MINIng), a flexible software package for exploring all forms of human genetic variation. Unlike existing tools, GEMINI integrates genetic variation with a diverse and flexible set of genome annotations (e.g., dbSNP, ENCODE, UCSC, ClinVar, KEGG) into a unified database to facilitate interpretation and data exploration. Whereas other methods provide an inflexible set of variant filters or variant prioritization methods, GEMINI allows researchers to compose complex queries based on sample genotypes, inheritance patterns, and both pre-installed and custom genome annotations. GEMINI also provides methods for ad hoc queries and data exploration, a simple programming interface for custom analyses that leverage the underlying database, and both command line and graphical tools for common analyses. We demonstrate GEMINI's utility for exploring variation in personal genomes and family based genetic studies, and illustrate its ability to scale to studies involving thousands of human samples. GEMINI is designed for reproducibility and flexibility and our goal is to provide researchers with a standard framework for medical genomics.

## Introduction

Unraveling the genetic components of human disease phenotypes demands not only accurate methods for discovering genetic variation, but also reliable strategies for interpreting the relevance of the identified variants. Owing to evermore accurate DNA sequencing technologies, human geneticists now have a potent tool for interrogating nearly every base pair in a human genome. Similarly, great algorithmic strides have been made[1,2] for identifying single-nucleotide and insertion-deletion polymorphisms from the billions of sequenced DNA fragments. However, given the scale and complexity of these variation catalogs and the formats that describe them[3], it remains a substantial challenge to manage and interpret genome-scale variation in the context of a disease phenotype. While itself limited, we best understand the consequences of genetic variation affecting protein-coding genes. Yet as recent studies of loss-of-function variation have shown, ostensibly damaging variants are frequently artifacts of data, annotation, or analysis[4,5].

As such, care must be taken in prioritizing potentially causal variants even in this seemingly "simple" case. Interpretation is far more challenging in the case of non-coding variation, as we have only a preliminary grasp of the functional consequences of non-coding variation on gene regulation and fitness[6-8]. Integrating functional genomics annotations from ambitious projects such as ENCODE will thus be crucial to assessing the impact of non-coding variation.

Given these analytical challenges, systematic efforts to identify genetic variation underlying disease phenotypes through exome and genome sequencing clearly depend upon the ability to query variants in the context of the incredible wealth of genome and "epigenome" annotations that have been curated since the completion of the human genome. The reality, however, is that this goal poses both technical and methodological challenges: genome annotation datasets are often quite large and are described in myriad file formats. Moreover, they come with varying documentation, they are frequently modified or updated, and they are housed in both centralized repositories[9,10] and on individual laboratory websites. Substantial technical ability is consequently required for even the most basic exploratory analysis integrating diverse genome annotations; greater analytic sophistication requires intricate, lab-specific pipelines that are laborious to produce and next to impossible to *reproduce*. We argue that the human genomics community needs flexible, reproducible, and scalable software for mining genome variation in the context of crucial genome annotations.

We have therefore developed GEMINI (GEnome MINIng), a novel software package that integrates genetic variation in the VCF format[3] with both automatically installed and researcher-defined genome annotations into a unified database framework. By integrating all forms of genetic variation (i.e., SNPs, INDELs, and structural variants) with diverse genome annotations, GEMINI allows both biologists and programmers to devise custom prioritization schemes for both coding and non-coding variants that meet their research criteria. The GEMINI database system provides a powerful variant analysis framework that eliminates the need to develop complex, and often error-prone, analysis pipelines.

GEMINI provides distinct functionality that is not available with existing software. Tools such as VEP[11] and snpEff[12] exclusively predict the consequence of variation on gene transcripts. Others such as ANNOVAR[13] and our own BEDTools[14] enable one to filter variants in a VCF file based on overlaps between a VCF file and individual annotations (e.g., a BED file of CpG islands). Integrating the many annotations necessary for genome-scale analyses in this manner inevitably requires custom scripts and is therefore laborious and error-prone. Other researchers have recognized this limitation and developed software that attempts to automate variant analysis and centralize genome annotations with genetic variants. However, extant software packages are either focused primarily on applying disease association tests to variants identified among a study cohort (e.g., PLINK/SEQ, unpublished), provide a limited set of annotations, or are more difficult to use because annotations are not directly integrated with genetic variation[15]. Moreover, few existing tools allow researchers to explore genetic variation with Structured Query Language (SQL), a powerful and expressive system for data analysis.

In contrast, as described in the following sections, GEMINI extends these basic data exploration concepts to allow researchers to: query variants and genome annotations in a common database framework using SQL, augment the database with custom annotations, prioritize variants based on sample genotypes and inheritance patterns, and develop powerful and reproducible analyses using a standard database framework and programming interface.

## Results

*Design overview.* As outlined in **Fig. 1**, GEMINI imports genetic variants and all sample genotypes from a VCF file into a SQLite (http://www.sqlite.org/) database (**Fig. 1A**). We prefer the use of a relational database to alternative, "NoSQL" approaches (e.g., Redis, MongoDB) because of the expressive power that SQL provides for constructing data exploration queries, its intuitive syntax, and its familiarity to many researchers. SQLite was chosen because of its speed, broad availability, and, in contrast to other database frameworks, its portability: a given GEMINI database can easily be shared as a single file among lab members and collaborators without a dedicated database server or additional configuration. Moreover, this portability allows researchers to "version" their research as samples and/or variant calling algorithms change by storing GEMINI databases along with the underlying VCF and sequence alignment files.

Each variant in an input VCF file is extensively annotated through automatic comparisons to a comprehensive and growing set of genomic annotation files including: dbSNP[16], ENCODE[6], ClinVar, 1000 Genomes[17], the Exome Sequencing Project[18], KEGG[19], and HPRD[20] (**Fig. 1B**; **Methods**). Annotated variants are loaded as rows in the `variants` database table. In the interest of reproducibility, the database also tracks (via the `resources` table) which version of the built-in annotations were used to create the database. Moreover, researchers may also augment the built-in annotations with custom annotation files relevant to their research (**Fig. 1C, Fig. 2A**). As we discuss in more detail below, storing extensively annotated variants in a relational database facilitates sophisticated data exploration via SQL queries and pre-defined GEMINI "tools" (a complete list of database tables and annotations are available in **Supplementary Table 1**). By using a database framework, we are able to not only index variants by their genomic coordinates, but also by their associated annotations. This expedites more sophisticated queries such as, "*what are all of the novel variants that overlap CpG islands and have an alternate allele frequency greater than 5% in my cohort?*" Such functionality distinguishes GEMINI from tools such as Tabix[21] and VCFtools[3] which can either index variants solely by genomic position, or isolate specific variants by scanning the entire VCF file (which are often tens or hundreds of gigabytes in size) for desired values in the VCF format's INFO field.

Studies of human disease require the ability to compare the genotypes of individual samples (e.g., cases versus controls) for each observed variant. A straightforward, yet impractical strategy for representing sample genotype information is to store the sample genotypes for each variant as distinct rows in a separate `genotypes` database table. In this model, accessing all observed genotypes for a given variant would thus require joining a `variants` table to a `genotypes` table, a strategy that scales very poorly when representing variation in VCF files with millions of variants and hundreds to thousands of samples. For example, merely one million variants for 1000 samples would yield 1 billion genotype rows and result in extremely poor query performance and scalability. Recognizing this limitation, we instead represent genotype information (genotype, phase, depth, etc.) for each sample as a compressed array that is stored as single column for each variant row (**Fig. 1D**). This inherently bounds the number of rows in the database to the number of variants observed. Moreover, since the proportion of rare variants will increase as a function of the number of samples, the majority of genotypes for rare variants will be identical and thus highly compressible. Therefore this strategy enables both query performance and scalability while still providing access to individual sample genotype information.

*Variant annotations.*
Discerning the functional relationship between experimentally identified genetic variants and a phenotype demands placing variants in the context of the extensive genome annotations that have been curated since the completion of the human genome. GEMINI integrates several commonly used genome annotations directly into the SQLite database including chromosomal cytobands, CpG islands, regions under evolutionary constraint, RepeatMasker[22] annotations, segmental duplications, known assembly gaps, mappability scores, and regional recombination rates (**Fig. 1B**).

In addition, several informative variant statistics and population metrics are calculated for each variant. The rationale behind this is that the VCF format is designed to store low-level sample genotype information such as the called genotype, its likelihood, and the sequencing depth that was observed for the sample. Consequently, it is often difficult to query VCF files based on summary genotype metrics such as the count of each genotype "type" (e.g., *how many heterozygotes were observed for this variant*?), or the count of samples lacking a called genotype. In an effort to facilitate downstream variant analysis, GEMINI derives and stores these and other metrics such as deviation from Hardy-Weinberg equilibrium, inbreeding coefficients, and nucleotide diversity estimates.

*Annotating coding variation.*
There are now several software packages[11-13] for predicting the impact of genetic variation on protein coding transcripts. Rather than reinvent the techniques already present in these tools, GEMINI currently integrates and standardizes predictions made by either snpEff or VEP. Each variant's clinical significance is also cataloged by comparisons to ClinVar (http://www.ncbi.nlm.nih.gov/clinvar/). In addition, GEMINI annotates functional pathway and protein-protein interactions through built-in KEGG and HPRD catalogs, thereby permitting researchers to explore the mutational burden in pathways and interacting proteins.

*Annotating non-coding variation.*
Assessing the consequence of non-coding variation remains challenging, yet new insights are being made by large-scale endeavors to map human regulatory elements among hundreds of cell types[6,8]. Nonetheless, attempting to understand non-coding variation in the context of disease requires the integration of many diverse genome annotations and exceedingly few tools exist to facilitate such research. As such, we have integrated three primary chromatin annotations from the ENCODE project: observed transcription factor binding sites[23], DNase1 hypersensitivity sites among 125 cell types[8], and Segway / ChromHMM consensus chromatin segmentation predictions among for 6 Tier 1 ENCODE cell types[24]. We anticipate continually extending and improving these annotations as dataset are made available from forthcoming efforts such as The Roadmap Epigenomics Project[25].

***The GEMINI database as a framework for data exploration and tool development.***
Our primary motivation for directly integrating genetic variants with genome annotations is to provide a flexible framework from which to explore genetic variation for disease and population genetic studies. Integrating these data in a single database provides a standardized and consistent interface for disease genetics, data querying and exploration, and new method development. Moreover, our design allows us to adapt to evolving research needs by quickly integrating new or improved genome annotations in order to facilitate analysis and future method development.

To demonstrate the analytic utility of the database framework, we provide several built-in tools for specific analyses (**Fig. 2A,B**). The *query* tool is arguably the most powerful as it allows the researcher to compose queries against the GEMINI database that satisfy their exact research question using both pre-installed and custom annotations. For example, **Fig. 2A** demonstrates a query identifying novel, rare (< 1% allele frequency), loss of function variants that meet an autosomal recessive inheritance model and overlap custom regions that are relevant to the researcher's disease of interest.

As illustrated in **Fig 1D,** sample genotype information (e.g., genotype, genotype "type" (heterozygote, homozygote, etc.), phase, and depth) is stored as database columns of compressed arrays. While this approach allows our database framework to easily scale to thousands of samples without generating billions of database rows, relational database systems do not inherently support queries that directly access individual genotypes. Since interrogating individual genotypes is crucial to studies of human disease, we developed an enhanced SQL interface that permits individual genotype queries (e.g., "`SELECT gts.proband, gts.mom, gts.dad`") and filters (e.g., "`gt_types.proband == HOM_ALT`").

In addition, we provide other tools that address more intricate research questions without requiring the researcher to write any analysis code (**Fig. 2B**). These include tools for identifying de novo mutations, as well as variants meeting both autosomal recessive and autosomal dominant inheritance patterns in family-based studies. In order to screen for these inheritance patterns, familial relationships must be defined in an optional PED file (**Fig. 1A**), which is subsequently stored in the "samples" table. We further provide methods for prioritizing loss-of-function variants and identifying putative compound heterozygous variants. By integrating pathway information from KEGG and protein interaction data from HPRD, we provide tools for exploring the functional pathways that variants affect, as well as networks of interacting proteins with multiple functional variants in a given sample.

Importantly, we enable researchers to augment the GEMINI database for their specific research needs. First, one may extend the database with genome annotations that are relevant to their own research. Secondly, researchers may create and integrate new analysis tools that leverage the GEMINI framework via Python scripts (**Supplemental Methods**). This flexibility will allow developers to extend GEMINI as new annotations and statistical methods (e.g., gene or region based burden tests) are developed (**Fig. 2C**). Moreover, recognizing that many researchers are uncomfortable with command-line data analysis tools, we have developed an interface for accessing the above tools and their results via a web browser. The browser interface integrates documentation of the database schema and the available tools, and connects directly with the IGV genome browser[26] allowing researchers to inspect the primary DNA sequence data underlying individual variants (**Fig. 3**).

*Performance*
Given the size and complexity of VCF files involving many samples, as well as the scale of genome annotation files, the time and resources required to import a VCF file and associated annotations into the GEMINI database were a fundamental concern in the design of the system we have developed. As such, loading a database can leverage multiple processors in order to enable reasonable database loading times. We support parallel processing on a single, multi-CPU machine and on common computing cluster frameworks (e.g., Sun Grid Engine, LSF, Torque): this inherent scalability will allow the framework to keep pace with future genetic studies involving thousands of samples. Moreover, parallel loading will enable the addition of new annotations without fearing dramatic reductions in performance. For example, using 4 processors, we were able to load a VCF file representing a twelve family member exome study (1.6 million variants; 12 genotypes per variant) in 41 minutes. Moreover, loading a VCF file including the genotypes of 1092 individuals from the 1000 Genomes Project (39.7 million variants; 1092 genotypes per variant) required 28 hours using 30 processors (**Methods**).

Since sample genotype information is stored as compressed binary arrays in the *variants* table and many annotations are stored more efficiently in a SQLite database than in a text-based VCF format, the resulting GEMINI databases require substantially less storage space than the original VCF files. For example, a compressed version of the above 1000 Genomes VCF file requires 144 gigabytes after annotation with snpEff. In contrast, the corresponding GEMINI database, with annotations, requires just over half the space (78 gigabytes). Moreover, the compression ratio only improves as the number of sample genotypes in the VCF file increases

## Discussion

We have developed a flexible new analysis framework that scales to the demands of both family-based disease studies and large-scale investigations involving thousands of individuals. By integrating genetic variation in the standard VCF format with a diverse and continually expanding set of genome annotations, GEMINI provides a uniquely powerful resource for exploring and interpreting both coding and non-coding genetic variation. Elucidating the genetic variants that underlie both unsolved single gene disorders and complex disease phenotypes requires the integration of a broad range of genome and disease annotations. Indeed a recent review of the challenges facing the interpretation of cancer genomes argues that a more detailed understanding of cancer etiology will require the integration of diverse information including pathway annotations, chromatin modifications, DNA methylation, and expression data[27]. GEMINI enables the integration of many large and heterogeneous genome annotations and as such, it provides a powerful tool that can address the analytical demands of complex disease research.

Given the clear necessity of such tools for advancing medicine in the genomic age, it is not surprising that several new commercial software packages have been developed in the last two years. Our goal is to provide a scalable, open-source medical genomics tool enabling other researchers to easily integrate new methods and genome annotations for the benefit of the human genomics research community.

## Methods

***Processing and loading VCF files***. Variant and genotype information stored in the VCF format is parsed and processed with the CyVCF Python module (http://github.com/arq5x/cyvcf). Each variant record in the input VCF file is screened for overlaps with genome annotation files using Tabix[21] and pysam[28]. Once all annotations have been collected for a variant, the variant record is inserted into the GEMINI SQLite database (in both the `variants` and the `variant_impacts` tables). When VCF files contain genotypes from many samples, simply reading and parsing the VCF file is time consuming. The additional cost of variant annotation causes the loading of a VCF file into GEMINI to be very computationally intensive. Therefore, in an effort to allow the loading to scale to the size of current and future VCF files including thousands of samples, the loading step can be parallelized on single machines with multiple CPUs. In addition, through use of the IPython.parallel library (http://ipython.org/ipython-doc/dev/parallel/) loading can be parallelized with computing clusters supporting LSF, Sun Grid Engine, or Torque load management systems.

***Annotation file management, provenance, and reproducibility***. Genome annotation files (e.g., UCSC or ENCODE tracks) can be easily added to the GEMINI framework, and since the loading step can be easily parallelized, the inclusion of new or updated annotation files in the interest of improving analyses will not substantially impact performance. We maintain an internal record of the annotation files used by a given GEMINI version and annotation files are stored on a public server in the interest of transparency. In addition, in support of research reproducibility, we document the provenance of each annotation file as well as any post-processing that was required to modify the files for use within GEMINI:
https://github.com/arq5x/gemini/tree/master/gemini/annotation_provenance.

***Storing and accessing sample genotype information***. Genotype information describing the genotypes, genotype types (e.g. 0 represents homozygous for the reference allele, 1 represents heterozygotes, etc.), genotype phase, and the aligned sequence depth leading to a sample's genotype is stored in the `gts`, `gt_types,` `gt_phases,` and `gt_depths` columns, respectively. Each column stores a compressed array in binary format, where each entry in the array represents the relevant genotype information for each sample. By compressing the arrays, we minimize the size of the database, since the genotypes for most samples will be identical, especially since most variants are rare. The GEMINI query framework provides an interface for accessing genotype information for individual samples by mapping sample names to the appropriate "bucket" in the compressed arrays stored in the database table.

*Implementation details*. The GEMINI software is implemented in Python and is available as a suite of command line tools and a web browser based interface. In addition, we maintain a Python programming interface that allows researchers to develop custom analysis scripts that leverage the GEMINI database framework in order to facilitate complex variant analyses.

## Software and Data availability
GEMINI is a freely available, open-source software package. The source code is maintained and available at: https://github.com/arq5x/gemini. Extensive documentation is available at: http://gemini.readthedocs.org/. GEMINI databases representing annotated variants from 1092 individuals from the 1000 Genomes project are available at: http://quinlanlab.cs.virginia.edu/.

## Acknowledgements
We are grateful for insightful comments and suggestions from Oliver Hofmann, Colby Chiang, Ira Hall, and Ryan Layer.

## Funding
This work was supported by an NIH award to AQ (NGHRI; 1R01HG006693-01).

## References

1    McKenna, A. *et al.* The Genome Analysis Toolkit: a MapReduce framework for analyzing next-generation DNA sequencing data. *Genome research* **20**, 1297-1303, doi:10.1101/gr.107524.110 (2010).
2    Li, H. *et al.* The Sequence Alignment/Map format and SAMtools. *Bioinformatics* **25**, 2078-2079, doi:10.1093/bioinformatics/btp352 (2009).
3    Danecek, P. *et al.* The variant call format and VCFtools. *Bioinformatics* **27**, 2156-2158, doi:10.1093/bioinformatics/btr330 (2011).
4    MacArthur, D. G. & Tyler-Smith, C. Loss-of-function variants in the genomes of healthy humans. *Hum Mol Genet* **19**, R125-130, doi:10.1093/hmg/ddq365 (2010).
5    MacArthur, D. G. *et al.* A systematic survey of loss-of-function variants in human protein-coding genes. *Science* **335**, 823-828, doi:10.1126/science.1215040 (2012).

6       Dunham, I. *et al.* An integrated encyclopedia of DNA elements in the human genome. *Nature* **489**, 57-74, doi:10.1038/nature11247 (2012).
7       Ernst, J. *et al.* Mapping and analysis of chromatin state dynamics in nine human cell types. *Nature* **473**, 43-49, doi:10.1038/nature09906 (2011).
8       Thurman, R. E. *et al.* The accessible chromatin landscape of the human genome. *Nature* **489**, 75-82, doi:10.1038/nature11232 (2012).
9       Kent, W. J. *et al.* The human genome browser at UCSC. *Genome research* **12**, 996-1006, doi:10.1101/gr.229102. Article published online before print in May 2002 (2002).
10      Hubbard, T. *et al.* The Ensembl genome database project. *Nucleic acids research* **30**, 38-41 (2002).
11      McLaren, W. *et al.* Deriving the consequences of genomic variants with the Ensembl API and SNP Effect Predictor. *Bioinformatics* **26**, 2069-2070, doi:10.1093/bioinformatics/btq330 (2010).
12      Cingolani, P. *et al.* A program for annotating and predicting the effects of single nucleotide polymorphisms, SnpEff: SNPs in the genome of Drosophila melanogaster strain w1118; iso-2; iso-3. *Fly* **6**, 80-92, doi:10.4161/fly.19695 (2012).
13      Wang, K., Li, M. & Hakonarson, H. ANNOVAR: functional annotation of genetic variants from high-throughput sequencing data. *Nucleic acids research* **38**, e164, doi:10.1093/nar/gkq603 (2010).
14      Quinlan, A. R. & Hall, I. M. BEDTools: a flexible suite of utilities for comparing genomic features. *Bioinformatics* **26**, 841-842, doi:10.1093/bioinformatics/btq033 (2010).
15      San Lucas, F. A., Wang, G., Scheet, P. & Peng, B. Integrated annotation and analysis of genetic variants from next-generation sequencing studies with variant tools. *Bioinformatics* **28**, 421-422, doi:10.1093/bioinformatics/btr667 (2012).
16      Sherry, S. T. *et al.* dbSNP: the NCBI database of genetic variation. *Nucleic Acids Res* **29**, 308-311. (2001).
17      Abecasis, G. R. *et al.* An integrated map of genetic variation from 1,092 human genomes. *Nature* **491**, 56-65, doi:10.1038/nature11632 (2012).
18      Tennessen, J. A. *et al.* Evolution and functional impact of rare coding variation from deep sequencing of human exomes. *Science* **337**, 64-69, doi:10.1126/science.1219240 (2012).
19      Kanehisa, M., Goto, S., Sato, Y., Furumichi, M. & Tanabe, M. KEGG for integration and interpretation of large-scale molecular data sets. *Nucleic acids research* **40**, D109-114, doi:10.1093/nar/gkr988 (2012).
20      Keshava Prasad, T. S. *et al.* Human Protein Reference Database--2009 update. *Nucleic acids research* **37**, D767-772, doi:10.1093/nar/gkn892 (2009).
21      Li, H. Tabix: fast retrieval of sequence features from generic TAB-delimited files. *Bioinformatics* **27**, 718-719, doi:10.1093/bioinformatics/btq671 (2011).
22      REPEATMASKER.
23      Neph, S. *et al.* An expansive human regulatory lexicon encoded in transcription factor footprints. *Nature* **489**, 83-90, doi:10.1038/nature11212 (2012).
24      Hoffman, M. M. *et al.* Integrative annotation of chromatin elements from ENCODE data. *Nucleic acids research* **41**, 827-841, doi:10.1093/nar/gks1284 (2013).
25      Bernstein, B. E. *et al.* The NIH Roadmap Epigenomics Mapping Consortium. *Nature biotechnology* **28**, 1045-1048, doi:10.1038/nbt1010-1045 (2010).
26      Robinson, J. T. *et al.* Integrative genomics viewer. *Nature biotechnology* **29**, 24-26, doi:10.1038/nbt.1754 (2011).
27      Garraway, L. A. & Lander, E. S. Lessons from the cancer genome. *Cell* **153**, 17-37, doi:10.1016/j.cell.2013.03.002 (2013).
28      PYSAM. http://code.google.com/p/pysam/.

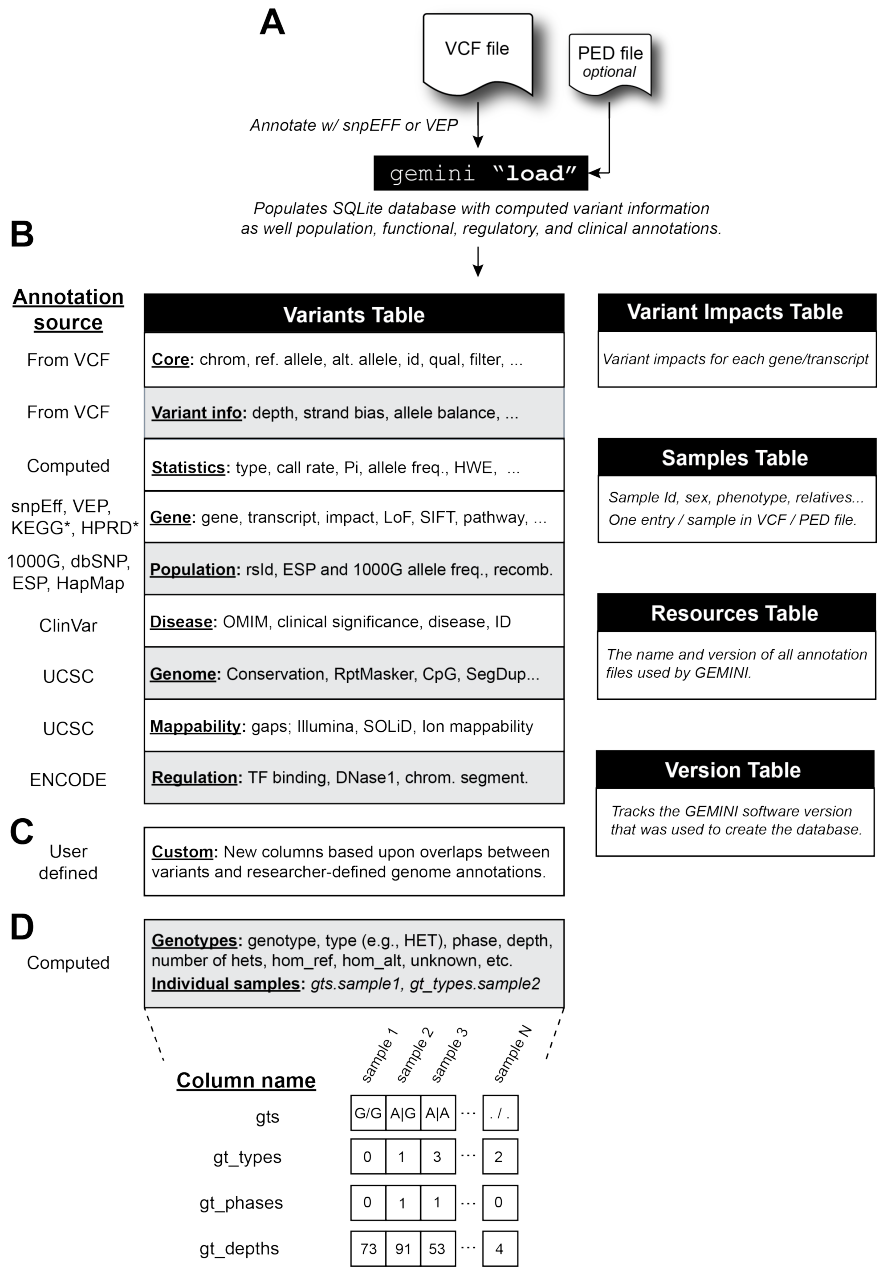

**Figure 1. Overview of GEMINI database and annotations. (A)** Genetic variants in the VCF format are loaded into the GEMINI database framework using the load sub-command. A PED file describing the sex, phenotype(s), and relatedness of the samples in the VCF may be provided to facilitate downstream analyses such as searches for de novo mutations or variants meeting specific inheritance patterns. **(B)** Each variant in the VCF file is annotated with information from several genome annotation sources that facilitate variant exploration and prioritization. The variants and associated annotations are stored in the variants and variant_impacts tables. **(C)** Researchers may also integrate their own annotations to facilitate custom analyses using annotations that are not pre-installed with the GEMINI software. **(D)** Genotype information for all samples is stored as a compressed array to enable database scalability and users may access genotype information for individual samples through an enhanced SQL interface.

(*) KEGG and HPRD annotations are not stored directly in the variants table, but are rather used in the context of specific GEMINI analysis tools.

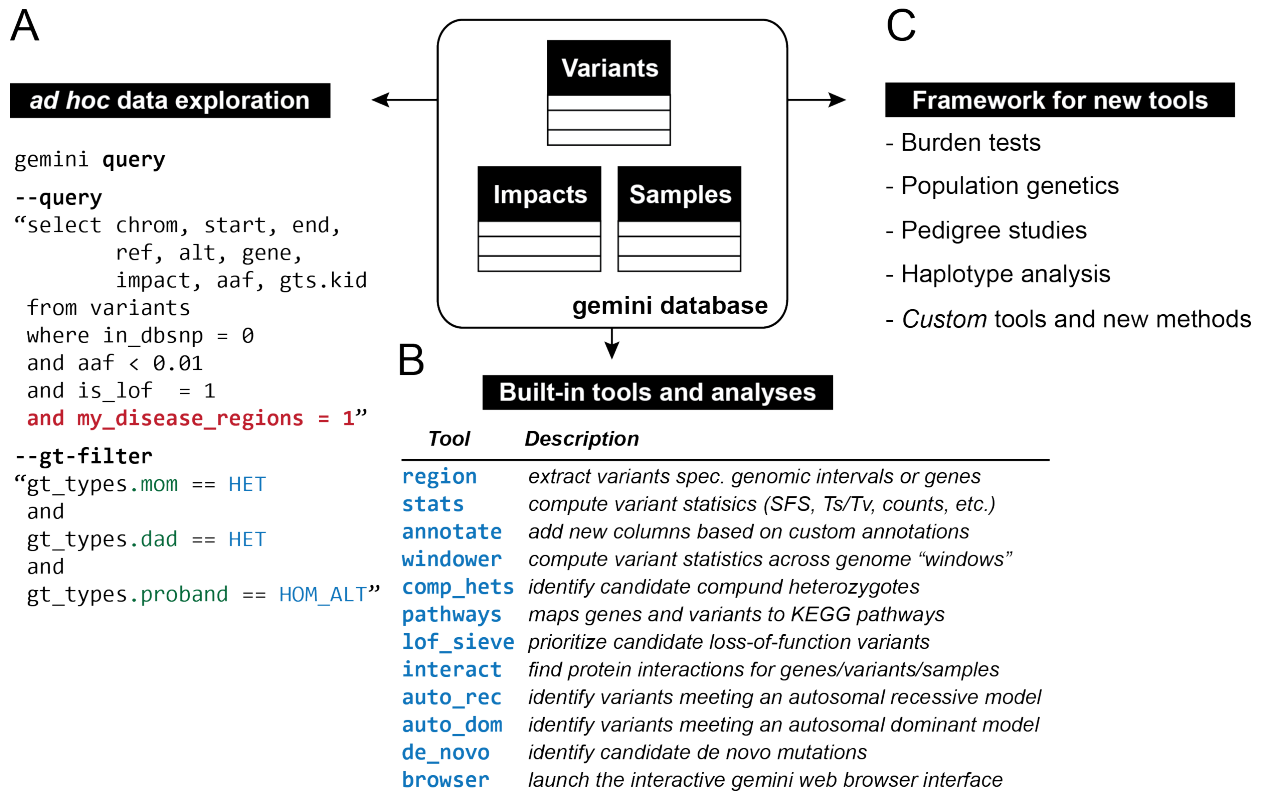

Figure 2. Variant mining and tool development with the GEMINI database framework. (**A**) Storing variants and annotations in the same database framework enable *ad hoc* SQL data exploration through both the `query` module and a Python programming interface. Analysis queries can filter variants based on pre-installed annotations (e.g., `in_dbsnp = 0`) and custom annotations (e.g., `my_disease_regions = 1`). Users may also select and filter variants based upon the genotypes of specific individuals (e.g., `gt_types.mom == HET`), thus allowing one to identify variants meeting specific inheritance patters, as shown here. (**B**) The GEMINI database framework also enables the development of tools that facilitate automated analyses for routine analysis tasks. (**C**) Moreover, the GEMINI database framework serves as a standard interface for developers to develop new tools and algorithms and to implement improved statistical tests for population and medical genetics.

**Figure 3. The GEMINI browser interface.** In an effort to enable collaborative research and to support users who are less comfortable working on a UNIX command line, we also provide a web browser interface to GEMINI databases. This figure depict the browser interface to the GEMINI `query` module; however, as illustrated in the navigation bar, interfaces also exist to other built-in analysis tools (e.g., for finding de novo mutations) and to the GEMINI documentation. **(A)** The browser interface to the `query` module allows users to run custom analysis queries in order to identify variants of interest. **(B)** Users may also enforce "genotype filters" that restrict the variants that are returned to those that meet specific genotype conditions or inheritance patterns. **(C)** Additional options are provided allowing the user to 1) add column headers describing the name of each column selected, 2) to create automatic links to the Integrative Genomics Viewer (IGV) from the reported variants, thus facilitating data exploration and validation, and 3) to report results to wither the web browser or to a text file for downstream analysis.